# Quantitative determination of interlayer electronic coupling at various critical points in bilayer MoS₂

**Authors:** Wei-Ting Hsu,[1,7][*] Jiamin Quan,[1][*] Chi-Ruei Pan,[2] Peng-Jen Chen,[2,3] Mei-Yin Chou,[2] Wen-Hao Chang,[4,5] Allan H MacDonald,[1] Xiaoqin Li,[1] Jung-Fu Lin,[6] and Chih-Kang Shih[1,7][+]

**Affiliations:**

[1] Department of Physics, The University of Texas at Austin, Austin, Texas 78712, USA

[2] Institute of Atomic and Molecular Sciences, Academia Sinica, Taipei 10617, Taiwan

[3] Institute of Physics, Academia Sinica, Taipei 11529, Taiwan

[4] Department of Electrophysics, National Yang Ming Chiao Tung University, Hsinchu 30010, Taiwan

[5] Research Center for Applied Sciences, Academia Sinica, Taipei 11529, Taiwan

[6] Department of Geological Sciences, Jackson School of Geosciences, The University of Texas at Austin, Austin, Texas 78712, USA

[7] Department of Physics, National Tsing Hua University, Hsinchu 30013, Taiwan

[*]These authors contributed equally to this work

[+]Correspondence should be addressed to: shih@physics.utexas.edu (C.K.S.)

**Abstract: Tailoring interlayer coupling has emerged as a powerful tool to tune the electronic structure of van der Waals (vdW) bilayers. One example is the usage of the "moiré pattern" to create controllable two-dimensional electronic superlattices through the configurational dependence of interlayer electronic couplings. This approach has led to some remarkable discoveries in twisted graphene bilayers, and transition metal dichalcogenide (TMD) homo- and hetero-bilayers. However, a largely unexplored factor is the interlayer distance, *d*, which can impact the interlayer coupling strength exponentially. In this letter, we quantitatively**

determine the coupling strengths as a function of interlayer spacing at various critical points of the Brillouin zone in bilayer $MoS_2$. The exponential dependence of the coupling parameter on the gap distance is demonstrated. Most significantly, we achieved a 280% enhancement of K-valley coupling strength with an 8% reduction of the vdW gap, pointing to a new strategy in designing a novel electronic system in vdW bilayers.

**Main Text**

**Introduction:** In the current topic of van der Waals hetero- and homo-bilayers, the interlayer electronic coupling plays a critical role in determining the electronic structures of the bilayer as a whole [1-7]. For example, in the subject of graphene twist-bilayer, the strength of interlayer coupling determines the magic angle that produces the novel Mott physics, superconductivity, and other strongly correlated electronic properties [8-12]. Similarly, in the transitional metal dichalcogenide (TMD) heterobilayers, the interlayer coupling determines the moiré potential landscape that gives rise to their tunable electronic and excitonic properties [11-18].

The strategy of moiré design is to harness the local stacking configurational dependence of the interlayer coupling. While the effect of the in-plane component of local stacking is well understood [19,20], the out-of-plane effect due to the variation of interlayer spacing is more difficult to capture quantitatively. One common approach is to use density functional theory (DFT) to calculate the configurational dependence of interlayer spacing and the electronic coupling strength; however, it is unclear if DFT can accurately describe such interlayer couplings. More significantly, there has been evidence that the variation in local interlayer spacing can be substantially different from the DFT result [6]. Thus, an independent experimental determination of the interlayer coupling as a function of interlayer spacing would play an important role in assessing the theoretical model in predicting the electronic structures of vdW hetero- and homo-bilayers. Moreover, such a quantitative determination will enable a new design parameter to tailor the electronic structure of vdW bilayers through interlayer spacing control, a strategy recently used for twisted graphene bilayers [21].

By using the Bernal-stacked bilayer $MoS_2$ as a model system, we report quantitative determination of interlayer coupling strength as a function of the interlayer spacing at different critical points of the Brillouin zone. The usage of Bernal stacked bilayer $MoS_2$ removes fabrication uncertainties associated with mechanical stacking of bilayers [18,22], making the data interpretation and extraction of coupling energy more straightforward. It also removes the lateral configuration variations in a moiré superlattice and allows one to assess the coupling as a function of layer spacing independently. By applying hydrostatic pressure in a diamond anvil cell up to 12.7 GPa, the interlayer spacing is changed from 0.62 nm to 0.57 nm, representing an 8% change. The coupling strength at various critical points is probed using a combination of differential reflectivity (DR) and photoluminescence (PL) spectra from which the exponential decay constants of the interlayer coupling at different critical points are determined. We further compare the results with the *ab-initio* calculations and find that all the results of K, Q and Γ points agree well with the DFT calculations. Importantly, after reducing the interlayer distance by 8%, the K-valley coupling strength is enhanced from 36 meV to 101 meV, representing a 280% enhancement. This result also points to a very promising strategy for designing novel quantum structures based on vdW bilayers.

**Results and discussion:** Conceptually, the interlayer electronic coupling in vdW bilayers can be described by a $2 \times 2$ Hamiltonian $H = \begin{bmatrix} \varepsilon_l & -t \\ -t & \varepsilon_u \end{bmatrix}$, where $\varepsilon_l$ ($\varepsilon_u$) is the single-particle energy level of the lower (upper) layer prior to coupling and $t$ is the coupling strength (also referred to as interlayer hopping integral). A larger coupling strength $t$ and/or a smaller energy difference ($\varepsilon_l - \varepsilon_u$) can lead to a larger energy splitting of the hybridized states. This model was successfully applied recently to capture the interlayer hybridization in commensurate hetero-bilayers [20]. Crucially, the coupling strength $t$ depends on several factors, i.e., the stacking configuration [19,20], the interlayer spacing $d$, and the critical point of the Brillouin zone. Due to the different projected atomic orbitals at different critical points, one expects to observe different coupling strengths at different valleys. Furthermore, when the interlayer spacing $d$ decreases, the increased orbital wavefunction overlap enhances the coupling strength $t$ and results in a larger energy splitting of the hybridized states, as depicted by the schematic shown in **Fig. 1(a)**.

In bilayer $MoS_2$, the effect of interlayer electronic coupling is manifested in the band splitting. **Figure 1(b)** shows the band structures for $MoS_2$ from 1 monolayer (black curve) to Bernal stacked bilayer (red curve) based on DFT at their natural state with $d = 0.62$ nm for the bilayer. **Figure 1(c)** shows the calculated band structure for the bilayer at a reduced interlayer spacing of $0.58$ nm (see **fig. S1** for other $d$). The critical points of interest are $K_V$, $Q_C$, and $\Gamma_V$. Note that interlayer coupling at $K_C$ is zero in the Bernal stacking due to symmetry [19,20]. An increase in interlayer coupling will lead to an increase in energy splitting at these critical points [23]. At $K_V$, the energy separation between $X_A$ and $X_B$ excitons in the DR spectrum can be used to determine the $d$-dependence of interlayer coupling [20]. The PL transition $X_I$ from the lower $Q_C$ state and the upper $\Gamma_V$ state (an indirect transition) can be used to determine the $d$-dependence of interlayer coupling at $Q_C$ and $\Gamma_V$. Two earlier studies by Dou *et al.* have reported the pressure-tuning of bandgap in bilayer $MoS_2$ up to $\sim 2 - 5$ GPa [24,25], indicating that interlayer coupling indeed depends on interlayer spacing. Here, we quantitatively determine coupling strength at various critical points, in which the magnitude and exponential dependence of $t$ are independently demonstrated. Our results are particularly useful for vdW bilayers consisting of monolayers with different chalcogen atoms such as $MoS_2/WSe_2$, where the lattice corrugation and/or structure reconstruction effects can result in a more than 20% modulation of the vdW gap [26].

Experimentally, we exfoliate the $MoS_2$ sample directly onto the diamond surface (**Figs. 1(d-e)**), where the monolayer, bilayer, and bulk regions can be identified by optical contrast and second harmonic generation measurements. The compressive pressure is uniformly transferred by the inert gas Ne medium. Due to the weak interlayer vdW force, the applied pressure leads to much greater compression along the out-of-plane direction [27], generating a quasi-vertical compressive strain on the 2D sample and resulting in a reduced interlayer spacing $d$. We use the lattice parameters obtained by X-ray diffraction (XRD) of thicker $MoS_2$ samples under pressure [27] to convert the pressure dependence to the interlayer spacing dependence.

Exciton resonances in $MoS_2$ monolayers and bilayers occur at the K valley and exhibit large oscillator strength. As shown in **Fig. 2(a)**, for a monolayer $MoS_2$, the spin splittings of the conduction band ($2\lambda_C$) and valence band ($2\lambda_V$) result from the spin-orbital coupling of the transition metal atoms [28-30]. The band splitting for bilayer and multilayer systems is further enhanced by the interlayer coupling, which is determined by the stacking configuration, band alignment, and valley spin [19,20]. For a Bernal-stacked bilayer (**Fig. 2(b)**), the coupling strength is zero (finite) for the conduction (valence) band edge [19,20]. In this context, the valence band splitting of the bilayer becomes $2\sqrt{\lambda_V^2 + t^2}$, which can be accessed optically. For example, two excitonic absorptions of K valley (known as $X_A$ and $X_B$ excitons) are typically observed in the DR spectra [31-33]. Since $2\lambda_v$ is much larger than $2\lambda_c$ [28-30], the energy difference between two bright excitons $\Delta E \equiv X_B - X_A$ becomes a good measure of the valence band splitting $2\lambda_v$ ($2\sqrt{\lambda_V^2 + t^2}$) for monolayer (bilayer) $MoS_2$ (also see **Note S1**).

**Figures 2(c)** and **2(d)** show the DR spectra of monolayer and bilayer $MoS_2$ with applied compressive pressure, respectively. While $\Delta E$ ($= 2\lambda_v$) of the monolayer is nearly independent of pressure, we find that $\Delta E$ ($= 2\sqrt{\lambda_V^2 + t^2}$) of the bilayer increases significantly with the applied pressure, demonstrating that $t$ is increasing. The energy evolution as a function of pressure is summarized in **Fig. 3(a)**. For the monolayer sample, we notice that there are rigid $X_A$ and $X_B$ peak blueshifts (~20 meV) between zero and 1.9 GPa, which may be caused by strain-induced changes in bandgap energy [27]. However, since our approach relies on energy difference, the influence of bandgap energy has been largely eliminated. Based on the model, we extracted the coupling strength $t^{(K_V)}$ as a function of pressure, as displayed in **Fig. 3(b)**. At the highest applied pressure (12.7 GPa), we determined a coupling strength $t^{(K_V)} = 101$ meV, which is ~280% greater than the value found at zero-pressure ($t^{(K_V)} = 36$ meV). We also observe that $t^{(K_V)}$ exhibits an increasing slope at larger pressures.

Here we use the lattice parameters obtained by XRD of thicker $MoS_2$ sample under pressure [27] to convert the pressure dependence to the $d$-dependence, as shown in **Fig. 3(c)**. In **Fig. 3(d)**, we compare the experimental results (dots) with the spacing-dependent $t$ extracted from the band structure calculations (blue line, DFT-1). In this calculation, we only change the interlayer spacing $d$ and keep the in-plane lattice as a constant. It is clear that the coupling strength $t^{(K_V)}$ exhibits an exponential growth as a function of $d$. The results are fitted by an exponential function $t^{(K_V)} = t_0^{(K_V)} e^{-(d-d_0)/\lambda_0^{(K_V)}}$, with the equilibrium spacing $d_0 = 0.6165\,\text{nm}$. We determine $t_0^{(K_V)} = 34\,\text{meV}$ (42 meV) and $\lambda_0^{(K_V)} = 0.046\,\text{nm}$ (0.070 nm) for the experimental (DFT-1) data. The discrepancy between the experimental and DFT results cannot be neglected, especially the decay length differs by about 34% (also see **Tables S1-S2**).

We realize that this difference arises from the effect of in-plane lattice compression, which occurs simultaneously when hydrostatic pressure is applied [27]. Based on the XRD data, we include both the in-plane and out-of-plane lattice compression in the calculation (red line, DFT-2). Clearly, after considering the effect of in-plane lattice compression, the calculations fit the experimental data much better. We obtain parameters of $t_0^{(K_V)} = 41$ meV and $\lambda_0^{(K_V)} = 0.054$ nm, which are only ~15% different from the experimental data. Therefore, our results demonstrate that both in-plane and out-of-plane lattice compression can enhance the interlayer coupling strength, with the latter playing a major role. The reason can be readily understood from the view of orbital wavefunction overlap: both in-plane and out-of-plane compression can bring the Mo atoms of adjacent layer closer, resulting in higher wavefunction overlap and greater coupling strength.

We further evaluate the coupling strength at other valleys using the pressure-dependent PL spectra of bilayer $MoS_2$ (**Fig. 4(a)**). The PL spectra feature two peaks, one from the direct K-K exciton and another from the indirect exciton $X_I$ [34,35]. In stark contrary to the nearly unchanged K-K exciton ($X_A$) energy, the indirect exciton $X_I$ exhibits a significant redshift of ~240 meV from zero to 6.3 GPa. The PL redshift indicates the shrinkage of the indirect gap under pressure. At pressure higher than 6.3 GPa, the $X_I$ indirect exciton emission is no longer detectable. After converting to the interlayer distance dependence, the indirect exciton transition energy as a function of interlayer distance, $X_I$ vs. *d*, is plotted in **Fig. 4(b)**. Also shown are DFT calculations for the $Q_C$-$\Gamma_V$ transition (red curve) and $K_C$-$\Gamma_V$ transition (blue curve) as a function of interlayer distance. Interestingly, the energy redshift of $X_I$ vs. *d* shows an excellent agreement with the $Q_C$-$\Gamma_V$ transition except for an apparent offset of ~0.25 eV. This offset is not surprising since DFT calculations often under-estimate the band gap [36]. Due to the close energy of $K_C$ and $Q_C$ points in the conduction band, the PL emission of the indirect gap in bilayer $MoS_2$ has been under debate [34,35]. Our results show that indirect PL emission is mainly from $Q_C$-$\Gamma_V$ excitons [24]. Under applying compressive pressure, the lifting-up of $\Gamma_V$ point and the pushing-down of $Q_C$ point both contribute to the shrinkage of the indirect gap (**Figs. 1(b)-(c)**).

Our optical spectroscopy measurements do not independently determine the interlayer coupling at $Q_C$ and $\Gamma_V$. Rather, it is a combined effect of $Q_C$ and $\Gamma_V$ as a function of *d*. However, given the fact the experimental result of $X_I$ vs. *d* shows excellent agreement with the calculated result for the $Q_C$-$\Gamma_V$ transition, we can conclude that the calculated results for the interlayer coupling at $Q_C$ and $\Gamma_V$ individually are accurate (also see **Tables S1-S2**). In this context, we are able to extract the coupling strength of $Q_C$ and $\Gamma_V$ points from the DFT calculations using the 2 × 2 Hamiltonian. As shown in **fig. S4**, the obtained results also exhibit an exponential dependence on *d* and can be fitted by an exponential function, with extracted $\lambda_0^{(\Gamma_V)} = 0.067$ nm ($\lambda_0^{(Q_C)} = 0.108$ nm) for $t^{(\Gamma_V)}$ ($t^{(Q_C)}$). When the pressure changes from zero to 6.3 GPa, we determine that $t^{(\Gamma_V)}$ changes from 335 meV to 522 meV and $t^{(Q_C)}$ from 192 meV to 253 meV.

In **Tables S1-S2**, we summarize and compare the experimental results with DFT calculations. We conclude that the coupling strength (decay length) of $\Gamma_V$ and $Q_C$ are larger (longer) than that of $K_V$. This valley-dependent coupling strength is due to the fact that the interlayer hopping integral is affected by the $p_z$ orbitals of the chalcogen atoms. Since the atomic distance between the inner chalcogen atoms of adjacent layers are the shortest, the large orbital wavefunction overlap leads to significant $p_z - p_z$ hybridization [37]. Our results show that the coupling strength is indeed correlated to the $p_z$-orbital component of the critical point, i.e., $p_z$-orbital components of $\Gamma_V$, $Q_C$ and $K_V$ points are $28\%(\Gamma_V) > 11\%(Q_C) > 0\%(K_V)$, which leads to $t_0^{(\Gamma_V)} > t_0^{(Q_C)} > t_0^{(K_V)}$. In addition, the decay length exhibits a trend of $\lambda_0^{(\Gamma_V)}, \lambda_0^{(Q_C)} > \lambda_0^{(K_V)}$, which can also be understood by $p_z - p_z$ hybridization. Since the wavefunction of $p_z$ orbitals elongates in the $z$-direction, it can be expected a less sensitive $d$-dependence, which leads to longer decay lengths at $\Gamma_V$ and $Q_C$ than at $K_V$. On the contrary, the states at $K_V$ comprise of transition metal $d_{x^2-y^2}, d_{xy}$ as the majority orbital and chalcogen $p_x, p_y$ as the minority orbital (see **Table S3**), leading to a rapidly decaying wavefunction along the $z$-direction and a shorter decay length.

In conclusion, our experiments firmly establish the exponential dependence of the electronic coupling on the interlayer spacing at various critical points of bilayer $MoS_2$. In particular, we quantitatively determine the coupling strengths of $\Gamma_V$, $Q_C$ and $K_V$ points, as well as their interlayer spacing dependence. We experimentally demonstrate that, due to the increased overlap of atomic orbital wavefunctions, interlayer coupling strength can be enhanced by both in-plane and out-of-plane lattice compression, with the latter playing a major role. By measuring the absorption of $X_A$ and $X_B$ excitons, $t^{(K_V)}$ and its $d$-dependence are independently determined, which are believed to be free from other factors such as band gap and exciton binding energy. In addition, we demonstrate that by applying a moderate pressure of 12.7 GPa, the interlayer K-valley coupling can achieve a 280% enhancement over that in natural bilayers. Through the PL measurement of indirect excitons, we also determine the $d$-dependence of $t^{(\Gamma_V)}$ and $t^{(Q_C)}$. These results have important implications for vdW heterostructures and moiré heterostructures. Tuning the magic-angle in twisted bilayer graphene through hydrostatic pressure has recently been demonstrated [21]. One therefore expects similar tunability can be achieved in twisted bilayer WSe$_2$ [11,12]. Most importantly, in the current topic of moiré designing of vdW hetero- and homo-bilayers, in which the shallow moiré potential depth has limited the observation of novel physics at low temperatures. A factor of three enhancement in modulation amplitude will liberate this limitation, thus profoundly impacting the applications of vdW bilayer system [38,39] in quantum information (e.g. quantum simulator) and other quantum phenomena.

## Materials and Methods

**Sample preparation.** High-quality single crystals MoS$_2$ (purchased from 2D semiconductors Inc.) were used in the experiments. First, the 2D layers were mechanically exfoliated on the polydimethylsiloxane (PDMS) substrates. For the pressure-tuning experiment, a pair of diamond anvils with a 400 μm culet was used [27]. A steel gasket was pre-indented and a hole of 220 μm diameter was drilled at the center of the pre-indented area. The MoS$_2$ monolayer and bilayer were then transferred to the sample chamber using a modified dry-transfer technique [40]. A ruby sphere was placed in the sample chamber and its fluorescence spectrum was used to calibrate the chamber pressure. Neon gas was used as the pressure-transmitting medium. Detailed pressure measurements can be found in **fig. S5**.

**Optical measurements.** Optical characterizations including PL, Raman, second harmonic generation (SHG), and DR spectroscopies were performed in a home-built optical microscope. The excitation source was focused on the sample surface by a 100× objective lens (N.A.= 0.5). The optical signal was then sent to a 0.5-m monochromator and detected with a nitrogen-cooled CCD camera. For PL and Raman measurements, a 532 nm solid-state laser (coherent Verdi v10) was used as the excitation source. For SHG measurements, the fundamental laser field was provided by a mode-locked Ti:sapphire laser with a wavelength of 880 nm. For DR measurements, the broadband white light source was provided by a fiber-coupled tungsten-halogen lamp. The integration time for each spectrum was 0.5-1 s, where the signal-to-noise ratio was further improved by averaging $> 100$ spectra.

**DFT calculations.** We performed first-principles calculations with density functional theory (DFT) using the Vienna *ab initio* simulation package (VASP) [41,42]. The interaction between valence electrons and ionic cores is handled by the projector augmented wave (PAW) method [43,44]. The exchange-correlation functional used is the Perdew-Burke-Ernzerhof (PBE) form within the generalized gradient approximation [45]. A vacuum region of about 7 Å (13 Å) is used to eliminate the spurious interaction in the slab calculation for single-layer (bilayer) $MoS_2$. The energy cutoff for plane waves is 600 eV, and the Brillouin zone sampling is done using a 12×12×1 Monkhorst-Pack grid for the structure relaxation [46]. The optB86b functional including the vdW correction is adopted to describe the non-local interaction between two layers of $MoS_2$ [47,48]. The atomic positions are fully relaxed until the force on each atom is smaller than $0.01 \; eV \cdot Å^{-1}$. The calculated lattice constants of monolayer and bilayer $MoS_2$ are 3.18 and 3.17 Å, respectively. The spin-orbit coupling effect is included in the electronic band structure calculations.

**Acknowledgments:** This research was primarily supported by the NSF Materials Research Science and Engineering Centers (MRSEC) under DMR-1720595. We also acknowledge support from the Welch Foundation (F-1672 and F-1662), the US NSF (DMR-1808751) and the U.S. Air Force (FA2386-18-1-4097). C.-R.P., P.-J.C., and M.-Y.C. acknowledge the support from Academia Sinica, Taiwan. W.-H.C. acknowledges the support from the Ministry of Science and Technology of Taiwan (MOST-110-2119-M-A49-001-MBK) and the support from the Center for Emergent Functional Matter Science (CEFMS) of NYCU supported by the Ministry of Education of Taiwan. W.-T.H. acknowledges the support from the Ministry of Science and Technology of Taiwan (MOST-110-2112-M-007-011-MY3) and the Yushan Young Scholar Program from the Ministry of Education of Taiwan. C.K.S. also acknowledge the Yushan Scholar Program from the Ministry of Education of Taiwan. We also thank Dr. Suyu Fu and Dr. Xianghai Meng for their help with the diamond anvil cells.

**Author contributions:** C.-K.S. and W.-T.H. conceived the idea and designed the experiment. W.-T.H. performed the spectroscopy measurements. W.-T.H., C.-K.S., and W.-H.C. analyzed the experimental data. The samples were exfoliated and stacked by W.-T.H. and J.Q., and supervised by X.L. J.-F.L. provided technical guidance for controlling interlayer distance with the diamond anvil cell. DFT calculations were performed by P.-J.C. and C.-R.P., and supervised by M.-Y.C. W.-T.H., W.-H.C. and C.-K.S. developed the model to interpret spectroscopic data. C.-K.S. and W.-T.H. wrote the paper with key inputs from A.H.M., M.-Y.C., W.-H.C., X.L. and J.-F.L. All authors discussed the results and commented on the manuscript.

**Data and materials availability:** All data needed to evaluate the conclusions in the paper are present in the paper and/or the Supplementary Materials. Additional data related to this paper may be requested from the authors.

**Competing interests:** The authors declare that they have no competing interests.

FIGURES

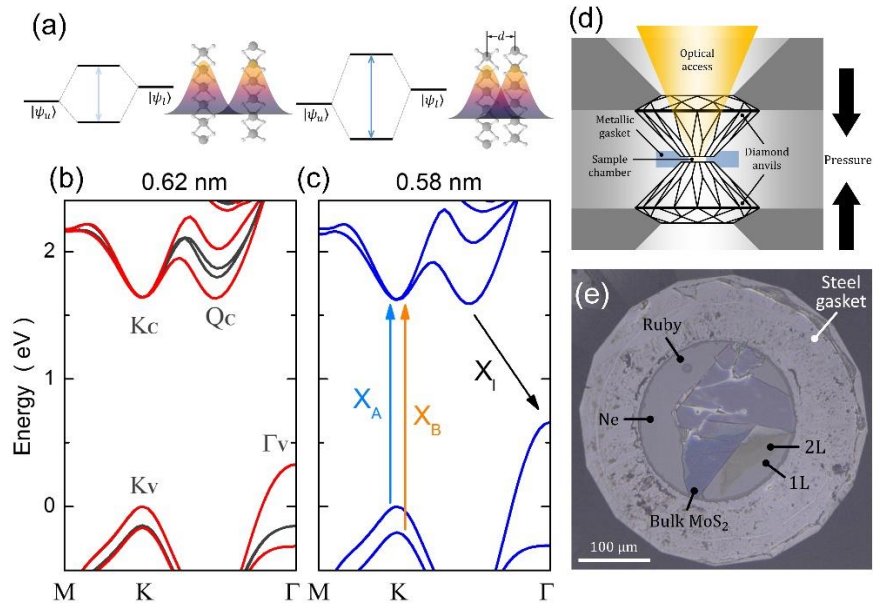

**Fig. 1. Control of interlayer electronic coupling using high-pressure diamond anvil cell.** (a) Schematic showing the interlayer hybridization of $|\psi_u\rangle$ and $|\psi_l\rangle$. When the interlayer spacing $d$ decreases, the orbital wavefunction overlap increases, which enhances the coupling strength and leads to larger energy splitting of the hybridized state. (b, c) DFT band structure of monolayer $MoS_2$ (black) and bilayer $MoS_2$ with interlayer spacing $d = 0.62$ nm (red) and 0.58 nm (blue). At $\Gamma_V$, $Q_C$ and $K_V$ points, band splitting becomes larger for bilayer with a smaller $d$. (d) Schematic showing a diamond anvil cell that can be accessed by optical spectroscopy. (e) An optical image showing the monolayer (1L), bilayer (2L), and bulk $MoS_2$ loaded in the high-pressure chamber. A ruby sphere is used as a pressure calibrant with the Ne medium loaded in the chamber.

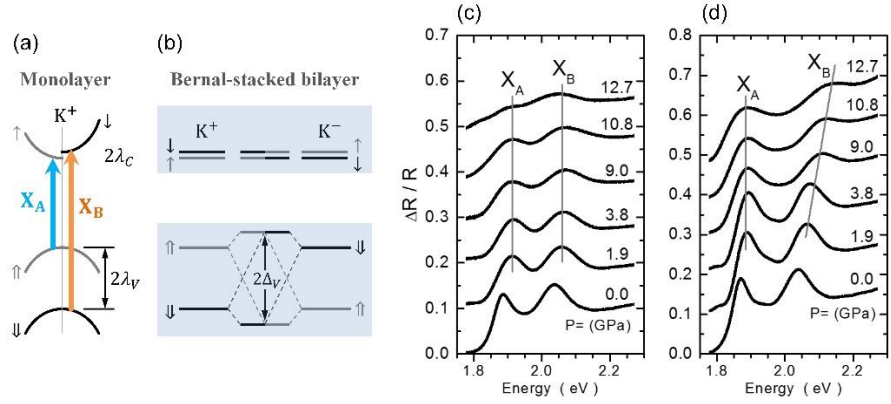

**Fig. 2. Pressure-tuning of interlayer hopping integral at K valley.** (a) Schematic showing the spin-allowed optical transitions ($X_A$ and $X_B$) at $K^+$ valley of monolayer $MoS_2$. The spin splitting of the conduction band ($2\lambda_C$) is much smaller than that of the valence band ($2\lambda_V$). (b) Schematic showing the interlayer hybridization of a Bernal-stacked bilayer, in which the valence band splitting is enhanced ($2\Delta_V > 2\lambda_V$). Note that the interlayer hopping is allowed (forbidden) at the valence (conduction) band edge. (c, d) DR spectra of the monolayer (c) and bilayer (d) $MoS_2$ under applied pressure. The spectra have been shifted vertically for clarity.

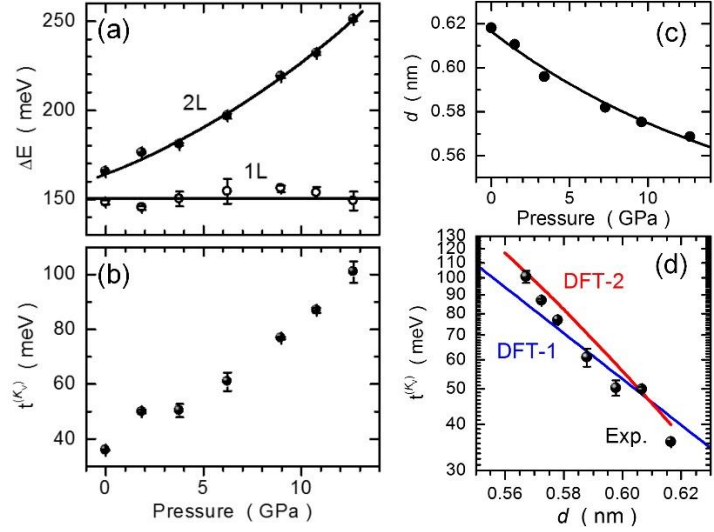

**Fig. 3. Determination of the spacing-dependent coupling strength.** (a) Pressure-dependent ΔE of monolayer and bilayer MoS$_2$. (b) Coupling strength $t$ as a function of applied pressure, showing a 280% enhancement at 12.7 GPa than that at zero GPa. (c) Pressure-dependent interlayer spacing $d$ determined by XRD. (d) Comparison of experimentally-determined (black dots) and DFT-calculated (blue/red curve) coupling strength as a function of $d$. In the calculation of DFT-1 (DFT-2), the compression of the in-plane lattice has not (has) been taken into account (see **figs. S1** and **S2** for band structures).

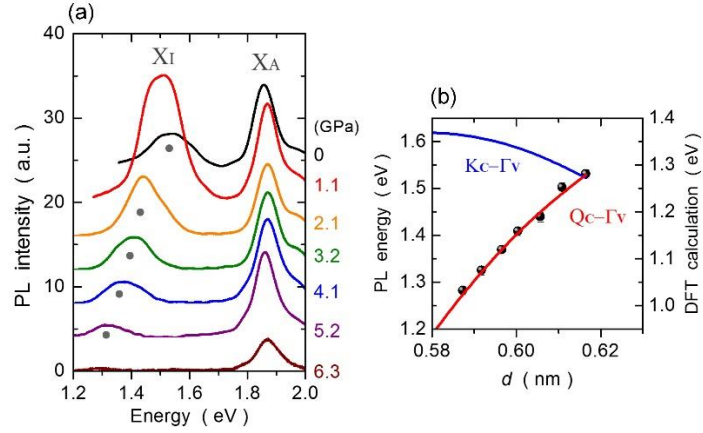

**Fig. 4. Pressure-tuning of the indirect gap.** (a) PL spectra of bilayer MoS$_2$ under applied compressive pressure. The indirect gap $X_I$ shows a significant redshift at high pressure. (b) A comparison of indirect gap measured by PL (dots) with the predicted evolution of E$_{Q\Gamma}$ and E$_{K\Gamma}$ gap, showing excellent agreement with the energy evolution of E$_{Q\Gamma}$ gap (red curve). The spacing dependence of $t^{(\Gamma v)}$ and $t^{(Qc)}$ extracted from DFT calculation can be found in **fig. S4**.